\documentstyle[multicol,aps,epsf]{revtex}
\begin{document}
\input psfig.tex
\draft

\title{Why Do Proteins Look Like Proteins?}
\author{Hao Li, Robert Helling\cite{hel}, Chao Tang, and Ned Wingreen}
\address{NEC Research Institute, 4 Independence Way, Princeton, New
Jersey 08540}

\date{March 5, 1996}
\maketitle

\begin{abstract}
Protein structures in nature often exhibit a high degree of regularity 
(secondary structures, tertiary symmetries, etc.) absent in random compact
conformations.  We demonstrate in a simple lattice model of protein
folding that structural regularities are related to high designability and
evolutionary stability.  We measure the designability of each compact
structure by the number of sequences which can design the structure, i.e.,
which possess the structure as their nondegenerate ground state.  We find
that compact structures are drastically different in terms of their
designability; highly designable structures emerge with a number of
associated sequences much larger than the average.  These structures are
found to have ``protein like'' secondary structure and even tertiary
symmetries.  In addition, they are also thermodynamically more stable than
ordinary structures.  These results suggest that protein structures are
selected because they are easy to design and stable against mutations, and
that such a selection simultaneously leads to thermodynamic stability.
\end{abstract}


\begin{multicols}{2}

Natural proteins fold into unique compact structures despite of the huge
number of possible configurations \cite{crei}.  It has been established
since Anfinsen that for most single domain proteins, the information coded
in the amino-acid sequence is sufficient to determine the three 
dimensional folded structure, which is the minimum free energy structure
\cite{anf}.  It is evident that protein sequences must be selected by 
nature such that they fold into unique three dimensional structures.  Since
folding maps sequences to structures, it is quite natural to ask whether
selection principles also apply to nature's choice of structures. Protein
structures often exhibit a high degree of regularity, e.g., rich secondary
structures ($\alpha$ helices, $\beta$ sheets) and sometimes even striking
tertiary symmetries, which are absent in random compact structures. What is
the origin of these regularities?  Does nature select special structures
for design?  What are the underlying principles governing the selection of
structures?

In this letter, we report our recent results from a simple model of protein
folding which suggest some answers to the above questions.  We focus on the
properties of each individual compact structure, by finding out the
sequences which have the given structure as their {\it non-degenerate}
ground state.  We show that the number of sequences $N_S$ associated with a
given structure $S$ differs drastically from structure to structure, and
that preferred structures emerge with $N_S$ much larger than the average.
These special structures are ``protein like'' with secondary structures and
symmetries, and are thermodynamically  more stable than ordinary structures.

Our results are derived from a minimal model of protein folding, which we
believe captures the essential ingredients of the problem. In this model,
a protein is represented by a self-avoiding chain of beads placed on a
discrete lattice, with two types of beads used to mimic polar (P) and
hydrophobic (H) amino acids \cite{dill}.  A sequence is specified  by a 
choice of monomer types at each position on the chain, $\{ \sigma_i\}$,
where $\sigma_i$ could be either H or P, and $i$ is a monomer index.  A
structure is specified by a set of coordinates for all the monomers 
$\{ {\bf r}_i\}$.  The energy of a sequence folded into a particular
structure is given by short range contact interactions,
\begin{equation}
H=\sum_{i<j}E_{\sigma_i\sigma_j}\Delta ( {\bf r}_i-{\bf r}_j),
\label{ham}
\end{equation}
where $\Delta ( {\bf r}_i-{\bf r}_j)=1$ if ${\bf r}_i$ and ${\bf r}_j$ 
are adjoining lattice sites but $i$ and $j$ are not adjacent in position 
along the sequence, and $\Delta ( {\bf r}_i-{\bf r}_j)=0$ otherwise.
Depending on the types of monomers in contact, the interaction energy will
be $E_{\rm HH}$, $E_{\rm HP}$, or $E_{\rm PP}$, corresponding to H-H, H-P,
or P-P contacts respectively (see Fig.2).

The above simple model has some justification in nature.  It is known that
the major driving force for protein folding is the hydrophobic force
\cite{kau}.  The tendency of amino acids to avoid water drives proteins to
fold into a compact shape with a hydrophobic core, and such a force is
effectively described by a short range contact interaction.  Although 
there are twenty different types of amino acids in nature, quantitative
analysis of real protein data reveals that they fall into two distinct 
groups (H or P) according to their affinities for water \cite{ltw}. 
There is also experimental evidence that certain proteins can be designed
by specifying only this HP pattern of the sequence \cite{hech}.

We choose the interaction parameters in Eq.~(\ref{ham}) to satisfy the
following physical constraints: 1) compact shapes have lower energies than
any non-compact shapes; 2) hydrophobic (H) monomers are buried as much as
possible, expressed by the relation $E_{\rm PP}>E_{\rm HP}>E_{\rm HH}$, 
which lowers the energy of configurations in which H's are hidden from 
water; 3)different types of monomers tends to segregate, expressed by 
$2 E_{\rm HP}>E_{\rm PP}+E_{\rm HH}$.  Conditions 2) and 3) were derived 
from our analysis of the real protein data contained in the 
Miyazawa-Jernigan matrix of inter-residue contact energies between 
different types of amino acids \cite{ltw}.

We have studied the model on a three dimensional cubic lattice and
on a two dimensional square lattice. We focus on the designability
of each compact structure.  Specifically, we count the number of
sequences $N_S$ which have a given compact structure $S$ as
their unique ground state.  This requires identification of the
minimum energy compact conformations of each sequence.  Since all 
compact structures have the same total number of contacts, we can freely 
shift and rescale the interaction energies, leaving only one free 
parameter.  Throughout this paper, we choose $E_{\rm HH}=-2.3$,
$E_{\rm HP}=-1$ and $E_{\rm PP}=0$ which satisfy conditions 2) and 3) above. 
The results are insensitive to the value of $E_{\rm HH}$ as long
as both these conditions are satisfied.
\begin{figure}
\narrowtext
\centerline{\epsfxsize=4.2in
\epsffile{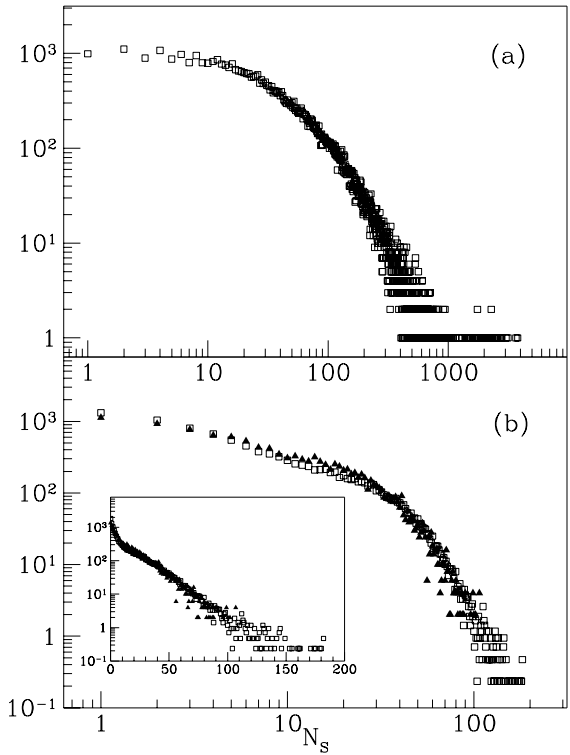}}
\vskip 0.2true cm
\caption{(a): Histogram of number of structures with a given
number of associated sequences $N_S$ for 3D $3\times 3\times 3$ case, in
a log-log plot. (b): Histogram of number of structures with a given $N_S$
for 2D $6\times 5$ (filled triangle) and $6\times 6$ (open square) case,
in a log-log plot. The bin size and rescaling of $y$ axis are
explained in the text. Insert: same data in a semi-log plot.}
\end{figure}

For the three dimensional case, we analyze a chain composed of 27 monomers. 
We consider all the structures which form a compact $3\times 3\times 3$
cube.  There are a total of 51704 such structures unrelated by rotational,
reflection, or reverse labeling symmetries.  For a given sequence,
the ground state structure is found by calculating the energies of all
compact structures.  We completely enumerate the ground states of all
$2^{27}$ possible sequences.  We find that $4.75\%$ of the sequences
have unique ground states.  As a result of this complete enumeration, we
obtain all possible sequences which ``design'' a given structure, i.e.,
have that structure as their unique ground state.  We denote by $N_S$ the 
number of sequences associated with a structure $S$.  In this way, the
number $N_S$ is a measure of the designability of a given structure, and
we have this information for all compact structures.

A surprising result is that compact structures differ drastically in
terms of their designability.  There are structures that can be designed
by an enormous number of sequences, and there are ``poor'' structures which
can only be designed by a few or even no sequences.  For example, the top
structure can be designed by $3794$ different sequences ($N_S=3794$), while
there are $4256$ structures for which $N_S=0$.  The number of structures
having a given $N_S$ decreases monotonically (with small fluctuations)
as $N_S$ increases (see Fig.~1(a)).  There is a long tail to the 
distribution.  Structures contributing to the tail of the distribution
have $N_S>>\overline {N_S}=61.72$, where $\overline {N_S}$ is the average
number.  We call these structures ``highly
designable'' structures.  The distribution is very different from the
Poisson distribution which would result if the compact structures were
statistically equivalent.  For a Poisson distribution with a mean
$\overline {N_S}=61.72$, the probability of finding even one structure
with $N_S>120$ is already $1.76\times 10^{-6}$. 

We observe that highly designable structures have certain secondary 
structures absent in random compact structures.  We examine the compact
structures with the ten largest $N_S$, and find that all have parallel
running lines folded in a regular way (see Fig.~2 for a typical example).
The number of straight lines (three amino acids in a row) found in 
these structures is 8 or 9, while the average structure has only 5.4
straight lines. 

To make sure that the above results are not artifacts of small size 
($3\times 3\times 3$ cube), we have also done systematic studies of size 
dependence in two dimensions (the study of larger structures in 3D is not 
practical due to limits of computing power).  We have studied systems of
sizes $4\times 4$, $5\times 5$, $6\times 5$, and $6\times 6$ on a 2D square
lattice.  For systems of sizes $6\times 5$ and $6\times 6$, a random 
sampling of sequences is performed \cite{note}.  To compare systems of
different sizes, appropriate rescaling of the axes is necessary.
We choose bin sizes for $N_S$ to be proportional to $\overline {N_S}$,
and rescale the number of structures by a factor proportional to 
the total number of structures. For the $6\times 5$ and
$6\times 6$ cases, to make sure that the random sampling of sequences
produces a reliable distribution, we double the number of sequences until
a fixed distribution is reached. 
 
We find that the  systems of different sizes in 2D all have the same
qualitative behavior as that found in 3D.  In each case, we find that
there are highly designable structures which stand out.  For the 
$6\times 5$ and $6\times 6$ systems where the total numbers of structures
are sufficiently large to produce smooth distributions, we find that the
two distributions have nearly identical shapes (see Fig.~1(b)).  We find
that the tail of the 2D distribution can be fitted by an exponential
function (see insert to Fig.~1(b)).  In contrast the tail in the 3D case
falls off slightly slower than  exponential.

Similar to the 3D case, 
we observe that the highly designable structures in 2D also
exhibit  secondary structures.
In the 2D $6\times 6$ case, as the surface to interior ratio
approaches that of real proteins, we find several interesting features.
Specifically, we find that the
highly designable structures often have
bundles of pleats and long
strands, reminiscent of $\alpha$ helices 
and $\beta$ strands in real proteins; in addition,
some of the highly designable structures have tertiary  symmetries (see 
Fig.~2 for a typical structure).
\begin{figure}
\narrowtext
\centerline{\epsfxsize=3in
\epsffile{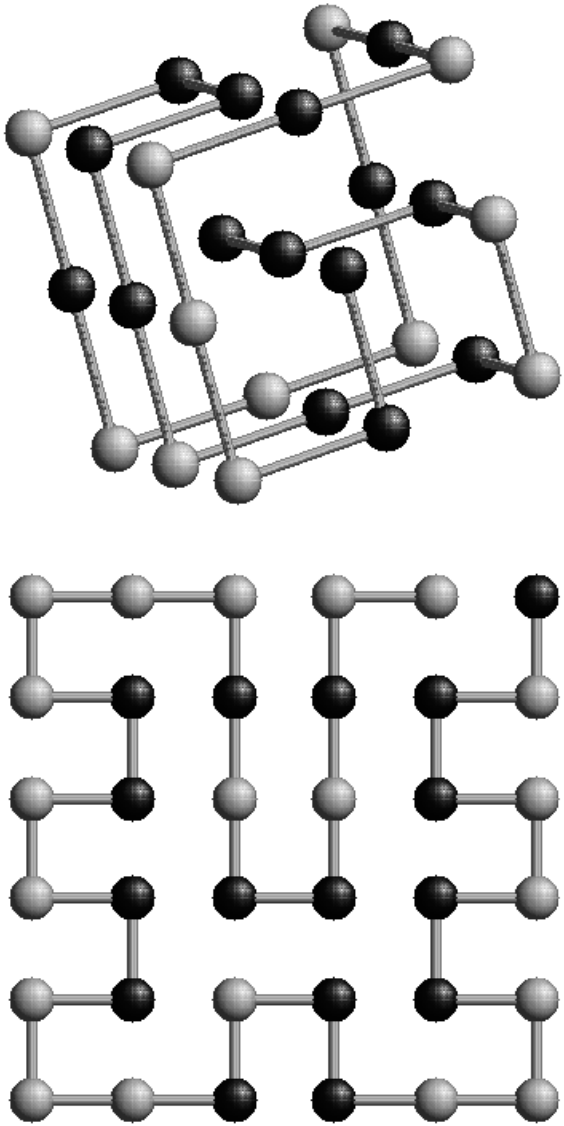}}
\vskip 0.1true cm
\caption{Structures with largest number of $N_S$ for 3D $3\times 3\times
3$ case (top) and 2D $6\times 6$ case (bottom). The sequences are one of
the $N_S$ possible sequences.  Beads colored black are of H type, and
beads colored light grey are of P type.  Two beads are considered to be
in contact if they are nearest neighbors but not connected by the
backbone.}
\end{figure}

A striking property of the highly designable structures is that they are,
on average, thermodynamically more stable than other structures.  The
stability of a structure can be characterized by the average energy gap 
$\overline {\delta_S}$, averaged over the $N_S$ sequences which design 
the structure.  For a given sequence, the energy gap $\delta_S$ is defined
as the minimum energy required to change the ground state structure to a
different compact structure.  For the 3D $3\times 3\times 3$ structures,
we find that there is a strong correlation between the number of sequences
$N_S$ and the average gap $\overline{\delta_S}$ (see Fig.~3). Highly
designable structures have average gaps much larger than those of
structures with small $N_S$, and there is a sudden jump in 
$\overline{\delta_S}$ for structures with $N_S\approx 1400$.
The number of structures with large gaps is 60.
The abrupt jump in  $\overline{\delta_S}$
is somewhat unexpected compared to the smooth distribution of
$N_S$. Such an abrupt transition provides a useful way of differentiating
the special, highly designable structures from the  ordinary ones.
According to this distinction, highly designable structures are only a
small fraction ($0.12\%$) of all the compact structures.
\begin{figure}
\narrowtext
\vskip -0.5true cm
\centerline{\epsfxsize=4.2in
\epsffile{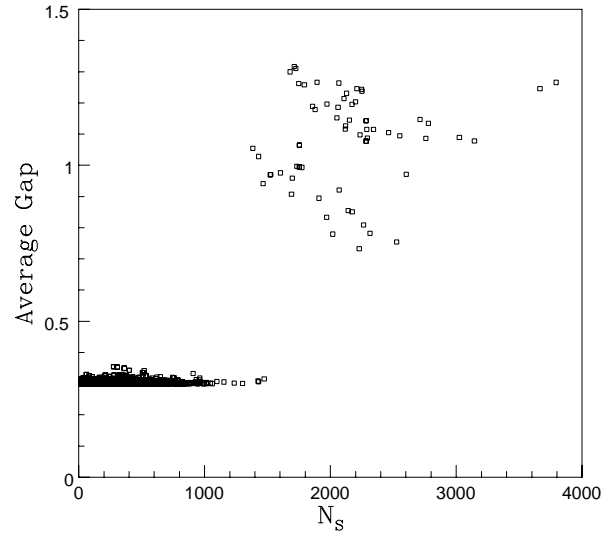}}
\vskip -0.5true cm
\caption{Average gap of 3D $3\times 3\times 3$ structures plotted
against $N_S$ of the structures.}
\end{figure}

The fact that highly designable structures are more stable than other
structures can be understood qualitatively in the following
way.  Consider a particular sequence associated with a highly designable
structure $S$.  A mutation of the sequence may change the energy of 
the structure $S$ as well as those of the competing structures.
If the gap is large, it is less probable that the energies of the
competing structures will shift below that of the  structure $S$.
Thus the structure $S$ is likely to stay as the ground state of the
mutant.  Therefore, a large gap is likely to correlate with a large 
number of sequences $N_S$ which design the structure.

\def\pp{{\cal P}_{\rm P}}
An important approach in studying real protein structures
is to study mutation effects and homologous sequences (sequences related
by a common ancestor in the past) \cite{albe}.
In our simple model, we call the $N_S$
different sequences that design the same structure ``homologous''. 
We have analyzed mutation patterns of the homologous
sequences for highly designable structures. The analysis
reveals phenomena similar to those observed in real proteins. For example,
we find sequences with no apparent similarities (with different types of
monomer at more than half of the sites) which can design the same
structure.  We also find some sites are highly mutable while some
sites are highly conserved.  The conserved sites for a given structure
are generally those sites with the smallest or largest
number of sides exposed to water.  Fig.~4 shows the
probability ${\cal P}_{\rm P}$ of finding a  P monomer at a particular
site, calculated  for the structure with largest $N_S$ for the 3D 
$3\times 3\times 3$ case and the 2D $6\times 6$ case (structures shown in
Fig.~2).  For the 3D case, we find sites which are perfectly 
conserved with  ${\cal P}_{\rm P}=0$ and ${\cal P}_{\rm P}=1$.
\begin{figure}
\narrowtext
\centerline{\epsfxsize=4.2in
\epsffile{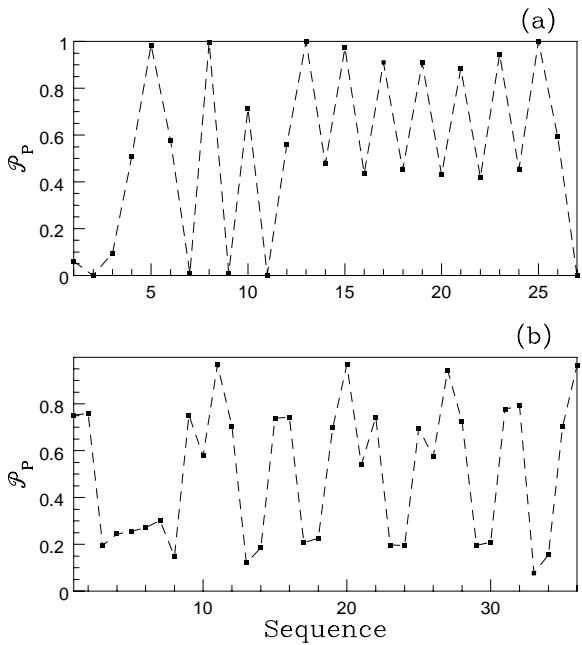}}
\vskip -0.2true cm
\caption{Probability of finding a site being a P type, calculated from
$N_S$ sequences of the top structures for 3D $3\times 3\times 3$ ((a)) 
and 2D $6\times 6$ ((b)) cases.}
\end{figure}

The mutation pattern for the 2D $6\times 6$ case shows additional
characteristics.  We observe that the pleated regions have alternating
arrangements (with period 2) of H and P types, the region of long folded
strands is essentially all H type, and the region connecting these
substructures (similar to turns in protein) contains primarily P type
monomers. 

Another way to characterize mutation patterns is to calculate the
entropy of the homologous sequences for a given structure
using ${\cal S}=\sum_{\rm sites}[-\pp\ln(\pp)+(\pp-1)\ln(1-\pp)]$. Given
only the knowledge of ${\cal S}$, an estimate of the number of all 
possible homologous sequences can be made, $N_{\rm est}=\exp({\cal S})$, 
assuming that mutation of each site is independent. 
We calculate $N_{\rm est}$ for all the structures and compare it
with the exact value $N_S$. For the highly designable structures,
$N_{\rm est}$ is a good order-of-magnitude  estimate for
$N_S$. For example, for the top structure in 3D $3\times 3\times 3$ 
case, $N_{\rm est}/N_S\approx 3.5$.  However, for the less designable
structures, $N_{\rm est}$ drastically overestimates $N_S$.
The large deviation starts at $N_S\approx 1400$, at the boundary
between large gap and small gap structures. This indicates
that for highly designable structures, the mutations are
roughly independent, while they are highly correlated for
other structures. 

Although our results for the 3D case were derived for small
structures ($3\times 3\times 3$), we believe similar results
hold for much larger structures. There is
evidence to this effect  from recent studies of design
on larger structures by Yue and Dill \cite{yue}
using  a similar model. In a few cases studied by Yue and Dill, 
they found that sequences with a small ground state 
degeneracy corresponded to structures with certain 
protein-like secondary structures and tertiary symmetries.
In light of our findings, we believe that such protein-like structures
are the highly designable structures with large $N_S$.  This 
interpretation is different from that of Yue and Dill, who suggest
that having minimal degeneracy is enough to produce protein-like
secondary structure and tertiary symmetries.  From our results, we know
that it is possible to find sequences to uniquely design even ``poor''
structures.  It is the requirement that {\it many} sequences design a 
particular structure which leads to protein-like secondary structures
and tertiary symmetries.

Although the detailed structures of real proteins are determined by many
factors, e.g., hydrogen bonding, shapes of the amino acids, etc.,
our results from the simple model suggest that there is a principle of
design and evolutionary stability which should play a crucial role in
the selection of protein
structures, i.e., real protein structures must be highly designable and
mutable.  Since highly designable structures are also more stable, such 
a selection principle solves the thermodynamic stability problem
simultaneously.  From an evolutionary point of view, highly designable
structures are more likely to be picked through random selection of 
sequences in the primordial age, and they are stable against mutations.

Our proposed principle of selection based on designability and mutability
should have important corollaries in protein structure prediction and
design.  If in fact nature only selects highly designable structures,
then structure prediction algorithms should limit the search of the 
conformational space to these special structures, which could be only a
tiny fraction of the total number of possible structures.  In fact,
a quite successful algorithm for structure prediction has been developed
recently, by using the templates from known protein structures \cite{jon}.
Our study lends theoretical support to such an approach.  Further
improvement depends on finding practical ways of identify highly
designable structures.

In conclusion, we find that there is a small fraction of compact 
structures which are highly designable and mutable.  These preferred
structures often have protein-like secondary structure and even tertiary
symmetries.  We find that highly designable structures are also more 
stable thermodynamically.  These results suggest that high designability
and evolutionary stability should play a crucial role in the selection
of protein structures, and that such a selection principle leads to 
thermodynamic stability at the same time.

An important question to ask is what is the kinetic accessibility
of these structures, and are there any other selection principles
imposed by the kinetics?  It is likely that the highly designable
structures are also easier to fold into, due to the large gap in their
excitation spectrum \cite{ssk}.  We have performed successful preliminary
folding simulations for some highly designable structures.  A more
systematic study of kinetics, including ordinary structures is underway.

We would like to acknowledge useful discussions with Bill Bialek, Michael
Hecht, Albert Libchaber, and Yigal Meir.

\end{multicols}

\end{document}